\documentclass[12pt]{article}

\usepackage{graphics}
\usepackage{amsmath}
\usepackage{amsfonts}

\newcommand{\T}{{\cal T}}
\def\eqn#1{$$#1$$}
\def\eqneqn#1{\begin{multline*}#1\end{multline*}}
\def\eqneqnn#1#2{\begin{multline}#2\label{#1}\end{multline}}
\def\eqnn#1#2{\begin{equation}#2\label{#1}\end{equation}}
\def\eln#1{\begin{align*}#1\end{align*}}
\def\elnn#1#2{\begin{align}\begin{split}#2\end{split}\label{#1}\end{align}}
\def\prt#1#2{\frac{\partial #1}{\partial #2}}

\def\xbar{ x \cos \omega t + y \sin \omega t}
\def\ybar{ y \cos \omega t - x \sin \omega t }
\def\xint{ \bar{x} \cos \omega t - \bar{y} \sin \omega t}
\def\yint{ \bar{y} \cos \omega t + \bar{x} \sin \omega t }

\begin{document}
\baselineskip=24 pt

\begin{center}
{\large The Energy of a Dynamical Wave-Emitting} \\
{\large System in General Relativity}
\\
F. I. Cooperstock and S. Tieu
 \\{\small \it Department of Physics and Astronomy, University 
of Victoria} \\
{\small \it P.O. Box 3055, Victoria, B.C. V8W 3P6 (Canada)}\\
{\small \it e-mail addresses: cooperstock@phys.uvic.ca, stieu@uvic.ca }
\end{center}
\begin{abstract}


The problem of energy and its localization in general relativity is critically 
re-examined. 
The Tolman energy integral for the Eddington spinning rod is analyzed in detail 
and evaluated apart from a single term. 
It is shown that a higher order iteration is required to find its value. Details 
of techniques to solve 
mathematically challenging problems of motion with powerful computing resources 
are provided. The next phase of following a system from static to dynamic to 
final quasi-static state is described.


\end{abstract}

PACS numbers: 04.20.Cv, 04.30.-w

\section{Introduction}\label{sect:introduction}

It has been said that ``a good problem shows its worth by fighting back". By 
that 
criterion, surely the problem of energy in general relativity is a very good one 
indeed: researchers continually return to it in spite of the fact that many 
might regard it as a ``solved" problem. An important part of the ``problem" 
aspect 
derives from the difficult issue of energy localizability in general relativity. 
Einstein \cite{einst}, Eddington \cite{edd} and others argued that energy is 
inherently non-localizable in general relativity while others such as Bondi 
\cite{bondinew} dismissed this notion. Through the years, various authors 
\cite{energyauth} \cite{alternative} have addressed the problem and recently, a 
new group of researchers have entered the debate \cite{energyauth2}.

While the localizability issue is of some interest even for static or stationary 
systems, of particular interest and importance is the issue in the case of a 
non-stationary gravity-wave-emitting system. This is due in part to the fact 
that although gravity waves have been researched intensively over many years, 
there is only sparse observational evidence, and that not actually direct, of 
the very existence of gravity waves. Moreover the existence of gravity waves 
carrying energy
would seemingly constitute a necessary condition for the viability of a quantum 
theory of gravity. Efforts to detect gravitational waves directly have emerged 
as a major research activity in recent years.

Various aspects of gravity waves led the first author \cite{coop} to the 
hypothesis that energy in general relativity is localized in the regions of the 
non-vanishing energy-momentum tensor $T_i^k$ (henceforth the ``localization 
hypothesis"). For example, plane fronted gravitational waves with parallel rays 
have been shown to be of the Kerr-Schild class and for the latter, it has been 
shown that all components of the pseudotensor vanish \textit{globally} 
\cite{gur}. This is in stark contrast to the truly tensorial corresponding case 
of plane 
electromagnetic waves that demonstrate a physically unambiguous flux of energy, 
both theoretically and observationally. Since there is a link between the Bondi 
news function and the pseudotensor \cite{coophob}, the same limitations would be 
drawn from this route of analysis. Another example is that of Bonnor's 
\cite{bondust} 
matching the aspherical Szekeres dust collapse metrics to a Schwarzschild 
exterior indicating a lack of energy flux. In this regard, it is also 
interesting to
recall remarks by Pirani \cite{pirani}: ``Suppose, for example that a 
Schwarzschild particle is 
disturbed from static spherical symmetry by an internal agency, radiates for 
some time, 
and finally is restored to static spherical symmetry. Is its total mass 
necessarily the same as before?"  
Very recent support for the localization hypothesis comes from the work of 
Bringley \cite{energyauth2} who has shown that Bonnor's directed beam of 
radiation is a Kerr-Schild metric.

Indications contrary to the localization hypothesis come from two directions: 
firstly, there is the gravitational geon construct, a bundle of gravitational waves that 
self-gravitate into a spherical ball with an exterior Schwarzschild metric 
indicating mass yet with a spacetime devoid of any energy-momentum tensor or 
singularities. However, we demonstrated that the time scale for the evolution of 
the geon was necessarily of the same order as the period of the constituent 
waves themselves, in contradiction with the assumptions entering into the 
construct of the geon itself; a perturbation analysis to test the requirement of quasi-stability results in a contradiction \cite{perrycoop}.
Secondly, there are the imploding/exploding Brill waves, again claimed to 
present mass without an energy-momentum tensor. However, this construct proceeds 
from an instant of time symmetry and it is at least questionable whether such an 
assumption is compatible with nonlinearly interacting waves. This second source 
of possible contradiction to the localization hypothesis will have to be 
considered in great detail.
In spite of all the years that have since passed and all the papers since 
written, the question remains.
As well, it is to be noted that the 
localization of energy in non-gravitational physics is in the region of non-
vanishing $T^{00}$. Hence the localization hypothesis may be viewed as a 
generalization of established physical phenomena and from this vantage point, 
not justifiably regarded as controversial.

If the localization hypothesis should prove to be 
correct, it would have fundamental consequences. First, it would imply that 
gravity waves in vacuum (assuming that they exist and there are ample reasons to 
believe that they do) would \textit{not} be carriers of energy, in conformity 
with the 
Kerr-Schild aspect. This notion challenges the very meaning that we give to the 
word ``wave", as a disturbance that \textit{carries energy}.  Second, without an 
energy aspect to the waves, it would be difficult to argue in favour of a 
quantum theory 
of gravity. In our earlier papers, we noted that while all particles and fields 
exist \textit{within} spacetime, gravity in essence \textit{is} spacetime 
itself, i.e. it is intrinsically different conceptually. Thus, it would not be 
surprising if the role of energy in general relativity should prove to be 
fundamentally different from that in other areas of physics.

Tolman \cite{tol} had found an expression for the total energy of a stationary 
system including the contribution from gravity as an integral over the region of the energy-momentum tensor. We had focused on this in earlier work \cite{roscoop} and more recently 
\cite{coopevidence} considered the time-rate of change of the Tolman integral at 
one instant as a vehicle to test the 
necessity condition for an energy loss via gravitational waves from the classic 
Eddington spinning rod \cite{edd}. The idea was that in the course of the 
evolution from initial stationarity to final (at least asymptotic) stationarity 
at both points at which the Tolman integral registers the energy, the Tolman 
integral should have undergone a change if energy were really being emitted. 
While the application of the time rate of change at one particular instant does 
not test with rigour the necessity condition for an energy loss for reasons that 
will be 
discussed later, it could possibly serve as a useful at least partial indicator 
as to how 
the gravitational contribution to the energy is being altered. At 
that time, the indications were that the Tolman integral was not changing. In 
this paper, we first derive a simpler expression for the time-rate of change   
of the Tolman integral. We then re-examine the Tolman integral more critically 
and find that the 
complete proper analysis 
is far more complex than originally envisaged. More information regarding the 
fields than that provided by Eddington and that we had used previously is 
required and this is furnished. In the 
course of the analysis, we focus upon the techniques of integration over the 
coordinates of the inertial frame and over the co-moving frame, using Dirac 
delta (a time-honoured helpful calculational tool) and step functions as required to properly delineate the source region and 
motion. These techniques will hopefully be found to be useful for researchers in 
the future. 

We find that all of the terms of the form field/$T_{ik}$ product cancel and we 
are left with 
three time derivatives of a trace quadrupole moment to evaluate, much as in 
traditional 
flux calculations which employ the untraced quadrupole expression \cite{ll} 
first derived 
by Einstein \cite{einst}.  However, for the latter,
the quadrupole terms are squared and hence are only required to low order of 
accuracy. By
contrast, in the present work, the quadrupole trace is not squared for the 
Tolman integral computation and
hence this quantity is required to higher order of accuracy, to a second 
iteration of 
the field equations. While this could possibly be examined in future research, 
it is valuable
to assess the broader picture. As Bondi, Bonnor, Feynman and others through the 
years had noted,
the truly reliable approach would be to examine the energy of a system at 
initial and final
stationary states of the system that has an intermediate wave-emitting phase. If 
there is a change,
then the existence and extent of an energy loss can be evaluated without 
ambiguity. The 
technique described above only examines the rate of change of the Tolman 
integral at an
instant of the dynamic phase. Actually, this rate should be integrated over the 
lifetime of the 
dynamic phase for a complete picture. There is an added motivation to do so when 
one considers
that the traditional sources studied have been assumed to be periodic as in the 
present problem,
and as Papapetrou showed many years ago, periodic sources are incompatible with 
the Einstein
field equations. Thus, the assumption of periodicity is at best an approximation 
which one
hopes is adequate when one undertakes analysis with the periodic assumption. 
Indeed if the system has developed over a
long period, then questions arise regarding the possibility that non-linearities 
could develop
to lead to a different field than is being envisaged. Moreover, the start-up and 
wind-down
phases of the motion are not being considered and they could have an important 
contribution.
To this end, we are presently in the process of analyzing a system with an 
explicit relatively
short evolution in time from a static to a dynamic to a quasi-static final 
state. This has the advantage
of providing the desirable total history as described and discussed above as 
well as mitigating any
onset of non-linear effects due to the relative brevity of the evolution.

For completeness, we review some of the fundamentals of the energy problem in 
general relativity and the work of Eddington and Tolman in 
Sec. \ref{sect:energyproblem}.
In Sec. \ref{sect:formalism}, the 
manner in which we set out to evaluate the time 
rate of change of the Tolman integral at a particular instant is developed and 
the framework for its application to the Eddington spinning rod is discussed. We 
illustrate the techniques that are employed to take time derivatives, comparing 
the manner of calculating in both co-moving and inertial frames. In Sec. 4, the 
details of the coordinate dependence for the Eddington problem are given. We 
supply the technical details that form the transition from the required time 
rate of change of the formal Tolman integral to the Maple program that is used 
to compute it
in Sec. 5, and we determine the time rate of change of the Tolman
integral up to one final term. In Sec. 6, we provide a summary and concluding 
discussions regarding 
the weak points of the assumptions leading to the previous results as well as a 
discriminating test and suggested future directions.

\section{The Energy Problem}\label{sect:energyproblem}

We recall that the generally covariant energy-momentum conservation laws 
required for general relativity \cite{conventions}
\begin{equation}\label{eq1}
	T_{i;k}^k = 0.
\end{equation}
can be cast into the form of an ordinary vanishing divergence
\begin{equation}\label{eq2}
	\frac{\partial}{\partial x^k}\left(\sqrt{-g}(T_i^k + t_i^k)\right) = 0
\end{equation}
by the introduction of the energy-momentum pseudotensor $t_i^k$ \cite{einst}. 
The pseudotensor  carries the specifically gravitational contribution to energy 
and momentum.  
It is a complicated expression involving the metric tensor and its first partial 
derivatives. Thus, global conservation laws can be extended into general 
relativity but at the expense of the inclusion of this non-tensorial object. As 
a result, the 
determination of a \textit{localized} expression for energy-momentum becomes 
problematic as the pseudotensor, unlike $T_i^k$, can be made to vanish at any 
pre-selected point by the proper choice of coordinates.

Einstein considered a weak perturbation of Minkowski space 
\begin{equation}\label{eq4}
	g_{ik}  =  \eta _{ik} + h_{ik}   
\end{equation}
in conjunction with the field equations
\begin{equation}\label{efeq}
	G_i^k= \frac{8 \pi G}{c^4}T_i^k.
\end{equation}
With the field perturbation $h_{ik}$, he computed the pseudotensorial Poynting vector flux to lowest non-vanishing order in $v/c$
from a bounded system emitting gravitational waves using the right hand side of  
(\ref{eq3}) below as
\begin{equation}\label{eq3}
	\frac{\partial}{\partial t}\int \left(\sqrt{-g}(T_0^0 + t_0^0)\right) dV
	= -c \oint\left(\sqrt{-g}t_0^\alpha \right) dS_\alpha
\end{equation}
where the integral on the left hand side was regarded as the energy E, extended 
throughout space via $t_0^0$.

An alternative approach is to proceed with a multipolar expansion of the field 
as in electromagnetism to express the rate of energy loss to lowest order as 
\begin{equation}\label{eq5}
	\dot{E} = -\frac{G}{45c^5}\left( \frac{d^3}{dt^3}
	\right) D_{\alpha \beta} \left( \frac{d^3}{d t^3}
	\right) D^{\alpha \beta}
\end{equation}    
where $D_{\alpha \beta}$ is the mass quadrupole tensor \cite{ll}. As applied to 
the Eddington spinning rod, both methods yield the energy loss rate
\begin{equation}\label{eq10}
	\frac{dE}{dt} = -\frac{32GI^2\omega^6}{5c^5}
\end{equation}
where $I$ is the moment of inertia, and $\omega$ is the angular velocity of the 
rigid (to lowest order) Eddington rod.

Later, Eddington \cite{edd} used a local approach analogous to that of radiation 
damping calculations in electromagnetism. With the conservation laws (\ref{eq1}) 
expressed as
(${\T}^{ab}$ is defined as $\sqrt{-g}T^{ab}$) 
\begin{equation} \label{eq6}
	\frac{\partial}{\partial x^k} \left( {\T}_i^k \right)
	= \frac{1}{2} {\T}^{ab} \frac{\partial h_{ab}}{\partial x^i},
\end{equation}
he integrated over a region just beyond the confines of the source and using 
Gauss' theorem, found a useful expression for the time rate of change of the 
integral of ${\T}_0^0$ as
\begin{equation} \label{eq7}
	\frac{\partial}{\partial t}\int{\T}_0^0 dV
	= \frac {1}{2} \int {\T}^{ab} 
	\frac{\partial}{\partial t} h_{ab} dV.
\end{equation}
The beauty of (\ref{eq7}) is that it enables one to focus upon the energy-
momentum tensor part of the energy expression without bringing in the non-
localized
pseudotensorial components as in (\ref{eq3}). Indeed , after an 
interesting sequence of argumentation in his book \cite{edd}, Eddington 
concluded that ${\T}^{00}$ was the appropriate density of energy in general 
relativity. Rather than the 
asymptotic field that is required in the flux calculation of (\ref{eq3}), the 
local field is required in calculations using the RHS of (\ref{eq7}). Eddington 
found this in the harmonic 
gauge as
\begin{equation}\label{eq8}
	h_{ab} = -4\int \left[ \frac{T'_{ab}
	-\frac{1}{2}\eta_{ab}T'}{r(1-\frac{v_r}{c}) }
	\right]_{ \text{ret}}dV'.
\end{equation}
To bring the elements of the retarded integral to a common time, Eddington 
performed a present time expansion as
\begin{equation}\label{eq9}
	\left[ \frac{T'_{ab}}{r\left( 1-\frac{v_r}{c}\right) }
	\right]_{ \text{ret}} = \frac{T_{ab}'}{r}-\frac{d}{dt}T_{ab}' 
	+ \sum_{n=2}^{\infty}\frac{(-1)^n}{n!}\frac{d^n}{dt^n}
	\left(r^{n-1}T'_{ab}\right).
\end{equation}
He considered the rod spinning in the $x-y$ plane and evaluated the right hand 
side of (\ref{eq7}) with the rod lying along the $x$ axis at $t=0$. When the 
three-volume element $dV$ is at $x$ and element $dV'$ is at $x'$, the distance 
between 
them at time $t$ is
\begin{equation}\label{eq7a}
	r=\sqrt{x^2 +{x'}^2 -2xx' \cos{\omega t}}
\end{equation}
which is used in conjunction with (\ref{eq8}), (\ref{eq9}) and (\ref{eq7}) and t 
is set to zero after the differentiations.

Simplifications occur because if $T^{ab}$ is any non-vanishing 
component at t = 0, ${T'}_{ab}$ will be an even function of time so odd order 
derivatives vanish and only the series

\elnn{eq7b}{
	\frac{\partial}{\partial t}
	\left[\frac{T'_{ab}}{ r (1-\frac{v_r}{c}) }\right]=&-
	\frac{d^2}{dt^2}{T'}_{ab}
	-\frac{1}{6}\frac{d^4}{dt^4}\left(T'_{ab}(x^2 +{x'}^2
	-2xx' \cos \omega t)\right)\\
	-&\frac{1}{120}\frac{d^6}{dt^6}\left(T'_{ab}{(x^2+{x'}^2
	-2xx'\cos{\omega t})}^2\right)+\cdots
}
is required.
With the rod spinning in the $x-y$ plane, the required transformation is

\elnn{trans}{
	\bar x^0 &\equiv \bar t = t, \\
	\bar x^1 &\equiv \bar x = x \cos \omega t + y \sin \omega t, \\
	\bar x^3 &\equiv \bar z = z \\
	\bar x^2 &\equiv \bar y = -x \sin \omega t + y \cos \omega t
}
where the $\bar x^i$-coordinates are co-moving with the rod and the 
$x^i$-coordinates are the inertial coordinates. This transformation is of 
sufficient 
accuracy for 
the calculations that we will require.

We will delve into greater detail than did Eddington. As the desired calculation 
is a perturbative one, we begin with the expression of the
energy-momentum tensor in the co-moving frame to the lowest order, 
\eln{
	\bar T^{00} &= \sigma \delta(\bar y), \\
	\bar T^{11} &= \sigma \frac{\omega^2}{2} (\bar x^2 - a^2 ) \delta(\bar y)
}
and all other components are zero. We can transform this into the
inertial coordinates using(\ref{trans}) to get
\elnn{Tintint}{
	T^{00} &= \bar{T}^{00} \\
	T^{01} &= -\omega y\bar{T}^{00} \\
	T^{02} &= \omega x\bar{T}^{00} \\
	T^{11} &= -\omega^2 y^2\bar{T}^{00} 
		+\cos^2\omega t \,\bar{T}^{11} \\
	T^{12} &=-\omega^2 xy\bar{T}^{00}
		+\cos\omega t\,\sin\omega t\,\bar{T}^{11} \\
	T^{22} &= \omega^2 x^2\bar{T}^{00} 
		+\sin^2\omega t\,\bar{T}^{11}.
}
Expressed in co-moving variables, they are
\elnn{Tintcomoving}{
	T^{00} &= \bar{T}^{00} \\
	T^{01} &= -\omega (\yint)\bar{T}^{00} \\
	T^{02} &= \omega (\xint)\bar{T}^{00} \\
	T^{11} &= -\omega^2 (\yint)^2\bar{T}^{00} 
		+\cos^2\omega t\, \bar{T}^{11} \\
	T^{12} &=-\omega^2 (\xint)(\yint)\bar{T}^{00}
		+\cos\omega t\,\sin\omega t\,\bar{T}^{11}\\
	T^{22} &= \omega^2 (\xint)^2\bar{T}^{00} 
		+\sin^2\omega t\, \bar{T}^{11}.
}
Eddington used the information he required from (\ref{Tintcomoving}) and again 
he found the result as in (\ref{eq10}). 

Eddington chose to regard the integral of ${\T}_0^0$ as the total energy of a 
system. In the past, we have referred to this integral as the material or 
``kinetic" energy
but this is not a truly adequate description because the root of the determinant 
of the total metric
tensor is included rather than that of the spatial part of the metric tensor 
that would provide proper 
three-volume. Thus, some field
contribution is included in the Eddington expression.
In later years, Tolman \cite{tol} (see also \cite{ll} for a more 
elegant derivation) showed that for stationary systems, the total energy is 
actually 
\begin{equation}\label{eq11}
	E=\int\left({\T}_0^0-{\T}_{\alpha}^{\alpha}\right) dV.
\end{equation}    
This is particularly attractive in that it is a wholly localized expression for 
bounded distributions and most importantly, it accounts fully for the 
contribution to 
the energy from the gravitational field.
Moreover, by comparison with (\ref{eq3}), we can identify the pseudotensorial 
gravitational 
contribution to the energy as deriving from the integral of the trace of the 
stresses/momentum flux densities. It was this portion of the gravitational 
contribution to the 
energy of a bounded stationary system that Eddington missed, his work pre-dating 
that of Tolman. We have seen its importance in various ways \cite{roscoop}. 
While we do not have a localized expression for energy in the non-stationary 
case, it is 
nevertheless of interest to ascertain whether or not this Tolman integral varies 
during a non-stationary phase. This is because it is the Tolman integral that 
will be the 
arbiter of energy change when the system eventually returns to stationarity. It 
must be stressed, however, that the information that we can glean at present is 
incomplete because we are incapable of measuring the change in the Tolman 
integral for the complete history stretching from the stationary start to the  
(asymptotically) stationary end. As well, there is the assumption of periodicity
with the dynamic field emerging from the source at a given instant without 
regard
to prior history. These aspects add further uncertainty.

There are alternative expressions for the mass of an isolated system. The Bondi 
mass \cite{bondi} m(u) is evaluated on the null cone. Since the transition to 
stationarity is an asymptotic one, the evaluation of the mass of system that is 
left behind after the dynamic period would entail the additional condition of 
allowing t to approach infinity using the Bondi expression, adding to the complexity.

Another choice is that due to Arnowitt, Deser and Misner (ADM) \cite{adm}. Their 
elegant 
formula expresses the mass of a stationary system as an asymptotic spacelike 
surface integral involving the asymptotic metric tensor components. We 
\cite{cooplim2} generalized the use of the ADM integral for non-stationary 
systems by treating it as a Poynting vector but we now see this as a 
pseudotensorial flux with the attendant ambiguities that this presents.

In our view, the Tolman integral is particularly attractive in that it is 
evaluated over the domain of the source and hence it yields the desired final 
state characteristic when the energy-momentum tensor no longer varies in time. 
This is because the evaluation of the Tolman mass at that point of the system 
that remains behind will be the same from that time onwards as the waves continue to 
infinity. What is being evaluated thereby is the end-state active gravitational 
mass of the material source including the contributions from its gravitational field 
and this is precisely what we seek in gauging whether or not there has been an 
energy loss of the system. One is not burdened by conditions on the asymptotic 
field in this approach.

\section{Formalism to Compute the Variation of the Tolman Integral}
\label{sect:formalism}

By raising an index, we can readily re-express the conservation law (\ref{eq6}) 
as \cite{coopevidence}
\begin{equation}\label{eq16}
	{\T}^{lk}_{,k}= F^l
\end{equation}
where
\begin{equation}\label{eq17}
	F^l=\frac{1}{2}{\T}^{ab}h_{ab,i}g^{il} +g^{il}_{,k}{\T}_i^k
\end{equation}
As before, in what follows, we will designate $x^i$ as the approximately 
inertial Cartesian 
coordinates. The Gauss theorem in conjunction with (\ref{eq16}) gives
\eqnn{eq19}{
	\frac{\partial}{\partial x^0}\int ({\T}^{\delta 0}x^\gamma
	+{\T}^{\gamma 0}x^\delta )dV
	=2\int{\T}^{\gamma \delta} dV+\int(F^\delta x^\gamma+F^\gamma x^\delta)dV
}
Similarly, we multiply by two $x's$ and integrate to get
\elnn{eq20}{
	\frac{\partial}{\partial x^0}\int {\T}^{00}x^\gamma x^\delta dV
	=\int({\T}^{0 \delta}x^\gamma+{\T}^{0 \gamma} x^\delta)dV 
	 +\int F^0 x^\gamma x^\delta dV.}
After taking $\frac{\partial}{\partial x^0}$ of (\ref{eq20}), eliminating
$\frac{\partial}{\partial x^0}\int ({\T}^{\gamma 0} x^\delta +{\T}^{\delta 0} 
x^\gamma ) dV$ by using (\ref{eq19}) and setting $\delta$ = $\gamma$ yields
\elnn{eq21}{
	\int {\T}^{\gamma \gamma} dV
	=& \frac{1}{2}\frac{\partial^2}{\partial (x^0)^2}\int {\T}^{00}
	x^\gamma x^\gamma dV-\int F^\gamma x^\gamma dV
	-\frac{1}{2}\frac{\partial}{\partial x^0}\int F^0
	x^\gamma x^\gamma dV\\.}
The integral of the spatial trace of the energy momentum tensor is required to 
complete the Tolman integral and its time rate of change is readily found as
\elnn{eq22}{
	\frac{\partial}{\partial x^0}\int{\T}_\gamma ^\gamma dV
	=\frac{\partial}{\partial x^0}\int{\T}^{\gamma k} g_{k \gamma}dV
	=&\int {T}^{\gamma \beta}_{,0}g_{\beta \gamma}dV 
	+\int{\T}^{\gamma 0}_{,0}g_{0 
	\gamma}dV+\int{\T}^{\gamma k}g_{k \gamma ,0}dV\\
	=&-\frac{\partial}{\partial x^0}\int{\T}^{\gamma \gamma}dV
	+\int{\T}^{\gamma \beta}_{,0}h_{\beta \gamma}dV
	+\int{\T}^{\gamma 0}_{,0}h_{0 \gamma}dV\\
	&+\int{\T}^{\gamma 0}h_{0 \gamma,0}dV +\int{\T}^{\gamma \beta}
	h_{\beta \gamma,0}dV}
using (\ref{eq4}) and $\eta_{\alpha \beta}$= diagonal$(-1,-1,-1)$.

The substitution of (\ref{eq21}) into (\ref{eq22}) yields
\elnn{eq23}{
	\frac{\partial}{\partial x^0}\int {\T}_\gamma^\gamma dV
	=& -
	\frac{1}{2}\frac{\partial^3}{\partial(x^0)^3}\int{\T}^{00}
	x^\alpha x^\alpha 
	dV+\frac{\partial}{\partial x^0}\int F^\alpha x^\alpha dV\\
	&+\frac{1}{2}\frac{\partial^2}{\partial(x^0)^2}\int F^0
	x^\alpha x^\alpha 
	dV \\
	&
	+\int{\T}^{\alpha k}_{,0}h_{\alpha k} dV
	+ \int{\T}^{\alpha k}h_{k \alpha ,0} dV
}
This is a simpler form of the required expression than we had found in earlier 
work (\cite{coopevidence}).
The second to fifth terms on the RHS of (\ref{eq23}) are of order 
$h_{ik}$ times $T_{ik}$ and hence of the familiar form of the RHS of the 
Eddington equation (\ref{eq7}). However the first term on the RHS of 
(\ref{eq23}) must be considered more carefully as there is no metric 
perturbation in it. While the untraced version of this term is only required to 
low 
order in the traditional flux calculations of the past, this is not the case in 
the present context
because we do not square this term to find the time rate of change of the Tolman 
integral. It is required to higher accuracy
and this will require a second iteration of the field equations.

It will also be of interest to evaluate the time rate of change of the 
``material" 
angular momentum \cite{cooprot}
\eqnn{rot1}{
   \frac{dL^{\gamma \delta}}{dt} = \frac{d}{dt}\int(x^{\gamma} {\T}^{0\delta}-
x^{\delta} {\T}^{0\gamma})dV.
}
Using (\ref{eq16}) and (\ref{eq17}) and Gauss' theorem, this becomes
\eqnn{rot2}{
   \frac{dL^{12}}{dt}=\int(F^2x^1-F^1x^2)dV
}
for the z component of angular momentum.

As an illustration of the techniques that we will employ, we compute the time 
derivatives of the mass quadrupole moment tensor  
\eqnn{quadrupole}{
	D_{\alpha\beta}
	= \int  \mu (3 x^\mu x^\nu \delta_{\alpha\mu} \delta_{\beta\nu}
	- \delta_{\alpha\beta} x^\mu x^\nu \delta_{\mu\nu} )\,dV
}
to the lowest order.

\begin{figure}

\begin{center}
\includegraphics{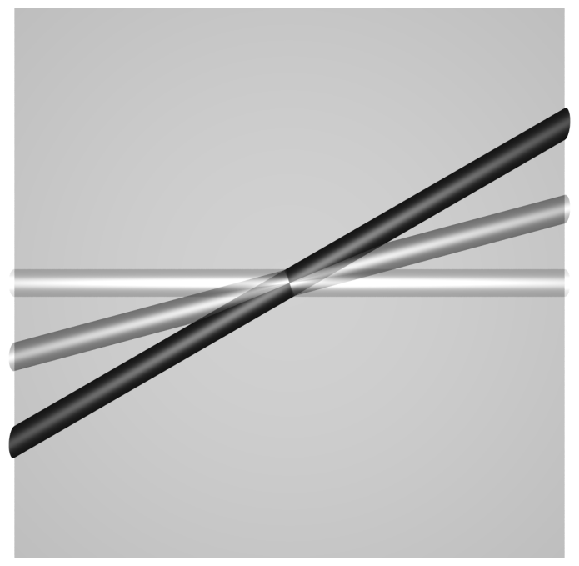}
\quad 
\includegraphics{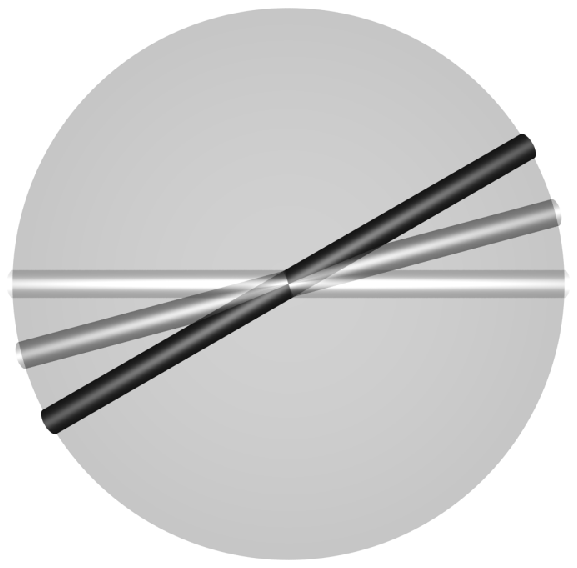}
\end{center}
\caption[Time-varying limits of integrations]{The figure on the left shows that
the limits of integration is $x \in[-a,a]$ whereas the figure on the right
shows the limits of integration as $x\in[-a\cos \omega t, a\cos \omega t]$.}
\label{fig:expandingrod}
\end{figure}

We can perform the calculation in the co-moving frame (as is commonly done with 
electromagnetic radiation analysis) or in the inertial reference frame. The 
latter has time-dependent
parts arising from the density
$\mu = \sigma \delta(z)\delta(\bar y)(H(a-\bar x)+H(a+\bar x)-1)$  where $H$ is 
the step-function.
Equation (\ref{quadrupole}) becomes
\eln{
	D_{\alpha\beta}
	= &\int_{-\epsilon}^{\epsilon}\int_{-\epsilon}^{\epsilon}\int_{-a}^{a}  
	\sigma \delta(z)\delta(\bar y)(H(a-\bar x)+H(a+\bar x)-1)
		(3 x^\alpha x^\beta 
		- \delta_{\alpha\beta} x^\mu x^\nu \delta_{\mu\nu} ) \,dx dy dz  \\
	= &\int_{-\epsilon}^{\epsilon}\int_{-\epsilon}^{\epsilon}
	\int_{-a\cos \omega t}^{a\cos \omega t}  
	\sigma \delta(z) \delta(-x \sin \omega t + y \cos \omega t)
	(3 x^\alpha x^\beta
	- \delta_{\alpha\beta} x^\mu x^\nu \delta_{\mu\nu} ) \,dx dy dz
}
for some $\epsilon>0$, large enough to enclose the Dirac-delta function.
The time-dependence of the limits of integration arises by virtue of the 
confinement of the source as viewed in the inertial frame. The rod is
confined by two factors, the Dirac-delta function, $\delta(\ybar)$, that 
maintains its position along the co-moving $\bar x$ axis 
and a double-step function, $(H(a-(\xbar))+H(a+(\xbar))-1)$, that specifies its 
length.  The double-step function forces the limits
of integration to be time-dependent.
Had we used fixed limits of integration, $x\in [-a,a]$
as opposed to $x\in [-a\cos \omega t, a\cos \omega t]$, it would have signified 
that the
rod was expanding as it sweeps through the $x$-axis as
shown in Figure \ref{fig:expandingrod}.
The rule for handling the $\delta(\bar y)dy$-integration,
as noted in the Appendix\footnote{The rule in
section \ref{appendix:diracdelta},
was aimed mainly at $t$ being in the the neighbourhood of $t=0$. However it
will also apply for $-\pi/2<\omega t<\pi/2$.  Furthermore, without loss of
generality, one can
argue that the calculation is similar for $\pi/2<\omega t<3\pi/2$, but with the 
$x$'s and $y$'s interchanging roles.
}, allows us to compute the following:
\eln{
	D_{11} &= \frac{2}{3} \sigma a^3 ( 3 \cos^2 \omega t  - 1 ),\\
	D_{12} &= \sigma a^3 \sin(2\omega t), \\
	D_{22} &= \frac{2}{3} \sigma a^3 ( 2-3 \cos^2 \omega t  ).
}
Applying this to (\ref{eq5}), we get the well-known
result of (\ref{eq10}), completely independent of time and hence without the
need to perform time-averaging.

Had we used fixed limits of integration,
we would have found that the resulting $dE/dt$ would have been {\em 
time-dependent}
with an incorrect value $dE/dt=-10 I^2 \omega^6$ at time $t=0$. (Furthermore, 
because of
a $\cos\omega t$ factor in the denominator, the time-average would have been
infinite as to be expected from an infinitely
expanding rod as $\omega t \rightarrow \pi/2$.) Thus, it is essential to bear in 
mind at all times that we are dealing with time-varying limits of integration.
While it would have appeared simpler to perform the calculations in the co-
moving frame as is commonly done using
\eln{
	D_{\alpha\beta}
	&= \int_{-\epsilon}^{\epsilon}\int_{-\epsilon}^{\epsilon}\int_{-a}^{a}  
	\sigma\delta(\bar{z})\delta(\bar{y})(3x^\alpha(t)x^\beta(t)
	- \delta_{\alpha\beta} x^\mu(t) x^\nu(t) \delta_{\mu\nu} )
	\, d\bar{x}d\bar{y}d\bar{z}
}
where $x(t)=\bar{x}\cos\omega t$ and $y(t)=\bar{x}\sin\omega t$, this would not 
serve our purposes in later work. The essential reason is that we will need to
take multiple derivatives in the {\em inertial} spatial directions.
Had we insisted on working in co-moving coordinates, we would have had to 
convert
back to the inertial coordinates to take the spatial derivatives and then 
convert forward to the
co-moving coordinates to perform the integration. As seen from the conversion
from (\ref{Tintint}) to (\ref{Tintcomoving}), this can be rather complicated.
Hence, we confine our calculations to the inertial coordinates.\footnote{The 
only exception to this is the integration of the source point
where $\bar{y}=0$.}

\section{The Extended Eddington Calculation}

Having established the limits of integration, we now re-examine the Eddington 
calculation (\ref{eq7}) as a second example. 
 Eddington had a one-dimensional problem involving only time differentiation
and hence was able to perform a simple line-integration.
However, the computation of the time derivative of the Tolman integral 
(\ref{eq11}) is considerably more complex. From the Eddington integral, 
(\ref{eq23}) must be subtracted and the latter, which contains $F^l$ as in 
(\ref{eq17}), has space as well as time derivatives.
Thus, one can no longer consider it a one-dimensional problem;
one must take spatial derivatives in directions parallel to and perpendicular
to the rod. Furthermore, the time-derivative of the
Dirac-delta distribution must be accounted for \footnote{When Eddington did his 
calculation in 1922,
the Dirac-delta distribution had not been established.} as well as the
double-step functions which truncates the rod. Clearly, there is a vast
requirement difference between Eddington's calculation and that of the Tolman
integral derivative.

The calculation is based on (\ref{eq4}) 
where $\eta_{ij}$=diag(1,-1,-1,-1) is the flat Minkowski metric and
\footnote{
 It is possible to use $h_{ij} = -4 \int_{-a}^a \int_{-a}^a \int_{-a}^a \Bigl[
	\frac{T'_{ij}-\frac{1}{2} \eta_{ij} {T'^m}_m}{
					\sqrt{ (x-x')^2+(y-y')^2 +(z-z')^2}}
	\Bigr]_{\text{ret}} \,dx' dy' dz' $
 and integrate in the inertial coordinate system but the integration would be
 extremely difficult; we choose to integrate with co-moving variables.
}
\eqn{
	h_{ij}(x,y,z,t) = -4 \int_{-\epsilon}^\epsilon\int_{-
\epsilon}^\epsilon\int_{-a}^a
	\Bigl[
	\frac{T'_{ij} - \frac{1}{2} \eta_{ij} {T'^m}_m  }{(1-v_r)r}
	\Bigr]_{\text{ret}} \,d\bar{x}'d\bar{y}'d\bar{z}'
}
where $r\equiv R(x,y;\bar{x}',t)
=\sqrt{ (x-\bar{x}'\cos\omega t)^2+(y-\bar{x}'\sin\omega t)^2+z^2}$.
Eddington used the present time expansion (\ref{eq9}).
This integration is performed in co-moving coordinates. The only place
that $\bar y'$ comes into play is in the $\delta(\bar y)$ and hence it is essentially zero after integration. 
For our purpose, we will not require the expansion beyond the 8$^{th}$ power.
The complete expression for $h_{ij}$ is
\eqneqn{
	h_{ij}(x,y,z,t) = -4 \int_{-\epsilon}^\epsilon\int_{-
\epsilon}^\epsilon\int_{-a}^a 
	\frac{T'_{ij}(\bar{x}',\bar{y}',\bar{z}',t)
		-\frac{1}{2}\eta_{ij} {T'^m}_m(\bar{x}',\bar{y}',\bar{z}',t)
		}{R(x,y,z;\bar{x}',t)}\\
	- \frac{d}{dt} \Bigl( T'_{ij}(\bar{x}',\bar{y}',\bar{z}',t)
		-\frac{1}{2}\eta_{ij} {T'^m}_m(\bar{x}',\bar{y}',\bar{z}',t)\Bigr)\\
	+\sum_{n=2}^\infty \frac{(-1)^n}{n!} \frac{d^n}{dt^n}
	\Bigl\{ R(x,y,z;\bar{x}',t)^{n-1} \Bigl(
	T'_{ij}(\bar{x}',\bar{y}',\bar{z}',t) 
		- \frac{1}{2} \eta_{ij} {T'^m}_m(\bar{x}',\bar{y}',\bar{z}',t)
	\Bigr) \Bigr\}
	\,d\bar{x}'d\bar{y}'d\bar{z}'.
}
It should be emphasized that
\eqn{
	T'_{ij}(\bar{x}',\bar{y}',\bar{z}',t)=T_{ij}(x,y,z,t)_{\bigl|
	x=\bar x \cos \omega t - \bar y \sin \omega t,
	y=\bar y \cos \omega t + \bar x \sin \omega t,
	z=\bar z
	}
}
is an inertial entity but it is expressed in co-moving variables.
Expressed in this form, there is no time-dependence in the Dirac-delta
or the double-step functions and hence the time-derivatives in the expansion
are straight-forward.

\section{Maple Calculations}

At this point, we supply the technical details that form the transition from the 
required time rate of change of the formal Tolman integral (the Eddington term 
minus (\ref{eq23})) to the Maple program that is used to compute it. 

\subsection{The Tolman Integrals}

The original Eddington calculation,
\eqn{
      \frac{d}{dt}\int \T_0^0 \,dV = \frac{1}{2}  \int T^{ij} \prt{h_{ij}}{t} 
\,dV
}
can easily be calculated using the following maple code
\begin{verbatim}
Eddington:= ... intxyz(1/2*TUPPER[i,j]*diff(hlower[i,j],t))
\end{verbatim}
The summations have been omitted for clarity. {\em The full Maple code is available from the authors.}

There are a few minor difficulties in calculating (\ref{eq23}).
As shown in the Appendix, the first time-derivative is commutative with
the time-dependent integral. Thus any terms having a single time derivative
can be readily computed without having to be concerned about boundary terms.
However, such boundary condition difficulties arising from higher derivatives do 
appear in the 
first and third terms of (\ref{eq23}).

We use the definition for $F^\alpha$ in (\ref{eq17}) to expand the
second term of (\ref{eq23}) to render it more manageable by Maple.
It is
\eqn{
	\prt{}{t} \int F^\alpha x^\alpha \,dV
	= -\int_{-\epsilon}^\epsilon\int_{-\epsilon}^\epsilon\int_a^a 
	\prt{}{t}\Bigl(
	\frac{1}{2} T^{ij} h_{ij,\alpha}
	-T^{ij} h^{k\alpha} \eta_{ik}
	\Bigr) x^\alpha \,dx dy dz
}
where  $\eta^{i\alpha}=-\delta^{i\alpha}$  has been used.  It is calculated 
using the following Maple codes: 
\begin{verbatim}
intxyz(diff(1/2*TUPPER[i,j]*diff(hlower[i,j],xi[alpha]),t)*xi[alpha])
intxyz(diff(-TUPPER[i,j]*diff(HUPPER[k,alpha],xi[j])
                                              *eta[i,k],t)*xi[alpha])
\end{verbatim}

The last terms are
\eqn{
	\prt{}{t} \int  T^{\alpha k} h_{\alpha k} \,dV
	= \int_a^a\int_{-\epsilon}^\epsilon\int_a^a 
	\prt{}{t}\Bigl( T^{\alpha k} h_{\alpha k}
	\Bigr) \,dx dy dz
}
with the equivalent code of
\begin{verbatim}
intxyz(diff(TUPPER[alpha,k]*hlower[alpha,k],t))
\end{verbatim}

For the third term, we require (\ref{secondderivativelemma}) because
of the occurrence of the double time-derivative.
\eqneqn{
	\frac{1}{2} \prt{^2}{t^2} \int F^0 x^\alpha x^\alpha \,dV  \\
	= -a\omega^2 (f(a)+f(-a))
	+\int_a^a\int_{-\epsilon}^\epsilon\int_a^a  \prt{^2}{t^2}\Bigl(
	\frac{1}{4} T^{ij} h_{ij,0} - \frac{1}{2} T^{ij} h^{k0}_{,j} \eta_{ik} 
	\Bigr) x^\alpha x^\alpha \,dx dy dz
}                                                                            
where the boundary terms are evaluated using the function,
\eqn{
	f(x)=\int_a^a \int_{-\epsilon}^\epsilon
	\Bigl(
	\frac{1}{4} T^{ij} h_{ij,0} - \frac{1}{2} T^{ij} h^{k0}_{,j} \eta_{ik} 
	\Bigr) x^\alpha x^\alpha \,dy dz.
}
This is calculated using the regular procedure, {\tt intxyz()}, and
the ``boundary'' procedure, {\tt intyz()}, as seen in the following 
code:
\begin{verbatim}
intxyz(diff(1/4*TUPPER[i,j]*diff(hlower[i,j],t)*(x^2+y^2+z^2)
  -1/2*TUPPER[i,j]*diff(HUPPER[k,0],xi[j])*eta[i,k]*(x^2+y^2+z^2),t$2))
intyz(1/4*TUPPER[i,j]*diff(hlower[i,j],t)*(x^2+y^2+z^2)
  -1/2*TUPPER[i,j]*diff(HUPPER[k,0],xi[j])*eta[i,k]*(x^2+y^2+z^2))
\end{verbatim}
The {\tt intyz()} lacks $x$-integration but evaluates the resulting functions
at $x=-a$ and $x=a$.  It also accounts for the $-a\omega^2$ factor as seen in the Appendix.

We find that \textit{apart from the quadrupole-like term, the second spatial 
trace part of the Tolman 
integral has a vanishing time-rate of change after all the remaining terms are 
summed}. 
In the absence of that single term, one would conclude that the Tolman integral 
changes at the same rate as the first part, i.e. as the Eddington value 
(\ref{eq7}), (\ref{eq10}). A goal in future work would be to evaluate this term 
by a 
further iteration of the field equations. However, this appears to be a very 
difficult task 
and the choice of a different dynamical system as we discuss below might obviate 
the necessity of a second iteration.

It is to be noted that by breaking up the left hand side of (\ref{eq7}) into a 
term that contains the determinant of the spatial part of the metric plus a term 
that carries 
the remainder which is of the form field$/{{\T}_0^0}$ product, we find that the 
time derivative 
of the remainder term is zero. Hence Eddington was actually calculating the time 
rate of change 
of what should be called the ``material" energy, i.e. that which arises from the 
proper 3-volume integral.

\subsection{The Angular-Momentum Integrals}
At this point, we gather additional information from the consideration of 
angular momentum change of the source.
Expanded, the expression for $dL^{12}/dt$ (\ref{rot2}) is
\eln{
	\frac{d L^{12}}{dt}
	&= \int x F^2 - y F^1 \,dV \\ 
	&=
	\int -\frac{x}{2} T^{ij} h_{ij,2} - x h^{i2}_{,j} T^{pj} \eta_{pi} \,dV
	- \int -\frac{y}{2} T^{ij} h_{ij,1} - y h^{i1}_{,j} T^{pj} \eta_{pi} \,dV
}
The equivalent codes are 
\begin{verbatim}
 intxyz(-x/2*TUPPER[i,j]*diff(hlower[i,j],y)
        -x*diff(HUPPER[i,2],xi[j])*TUPPER[p,j]*eta[p,i])
-intxyz(-y/2*TUPPER[i,j]*diff(hlower[i,j],x)
        -y*diff(HUPPER[i,1],xi[j])*TUPPER[p,j]*eta[p,i])
\end{verbatim}

When this is evaluated, we find that 
\eqnn{rot3}{
\frac{dL^{12}}{dt}= -\frac{32I^2\omega^5}{5}
}

Again, it is worth noting that a split of the right hand side of the above 
equation as we 
discussed in the case of the Eddington integral, leads to the analogous result: 
the calculation 
is the same as for the time rate of change of what we should rightfully call the 
``material" angular momentum.

Thus we see that $\frac{dE}{dt}$=$\omega\frac{dL}{dt}$ as expected for a rigid 
rod ( $ \frac{dI}{dt} = 0$) with kinetic energy $E$=$\frac{I\omega^2}{2}$ and 
angular momentum 
$L$=$I\omega$. We had also noted this result in our very early work that was 
based upon 
pseudotensorial fluxes. However, this relation arises from the classical kinetic 
energy and 
angular momentum connections with mass, moment of inertia and velocities. It is 
unclear 
whether or not this lowest order connection is appropriate to make the 
deductions regarding 
essentially higher order quantities and effects.

In addition, there are important caveats that we consider in the following 
section.

\section{Summary and Concluding Discussions}\label{sect:summary}

We began by tracing the essential features that have made the issue of energy 
and its localization in general relativity so challenging. The work of Einstein 
and Eddington were emphasized and as in earlier papers, we focused upon the 
problem of the Eddington spinning rod. An important attractive feature of this 
source of gravity waves is that to lowest order, it can be taken as rigid and 
hence the time variation derives solely from its variation in position. We noted 
that while Eddington considered the integral of ${\T}^{00}$ as the energy, some 
years later Tolman showed that for stationary systems, the energy is actually 
the integral of ${\T}_0^0-{\T}_{\alpha}^{\alpha}$ where the second term 
accounts for contributions to the energy stemming from the presence of
the gravitational field. Since 
the issue under consideration is that of the possible  \textit{gravitational} 
energy loss and since, if it exists, it is so incredibly minute in magnitude at 
least for the 
kind of sources that we are considering here, it is clear that this 
${\T}_{\alpha}^{\alpha}$ term is of considerable interest.

In earlier work, we had noted that various researchers through the years had 
presented the change in the energy as calculated from an initial stationary 
state to 
a final (at least asymptotically) stationary state as the ultimate criterion for 
the existence and measure of an energy loss. We then argued that a necessary 
condition for an energy loss was therefore a realization of a variation of the 
Tolman integral during an intervening non-stationary period of motion. We had 
previously used the field information supplied by Eddington to compute the 
change in the Tolman integral at an instant but this was inadequate. In this 
paper, we presented a complete and detailed treatment of the fields and found 
that the combination of all of the field/$T_{ik}$ product terms produced no net 
time variation
of the spatial trace part of the Tolman integral. The final term in the equation 
giving 
the time rate of change of this trace part of the Tolman formula is precisely 
the trace 
of the quadrupole moment expression that is used untraced in linearized theory 
with pseudotensor
to deduce the ``quadrupole formula". However, in the latter treatment, the 
expression
is squared and hence its value is required only to lowest order. In the present 
treatment
with the Tolman integral, it is not squared and the expression is required to 
sixth
order in $v/c$. While one can impose the condition of rigidity to lowest order 
as
Eddington had done, the field equations impose constraints on what is possible 
for
the energy-momentum tensor at higher order and the properties of the material 
come into
consideration. We faced this issue many years ago \cite{cooplim} for the problem 
of 
axially-symmetric free-fall using the pseudotensor and now we see the issue of 
structure appearing anew.
The exploration of this issue in the present context presents a new and likely 
difficult challenge that we leave for future study.

As well, there are other issues to be considered:

1. \textit{Is it sensible to be speaking of a localized energy in the first 
place?}

  Various researchers through the years have denied the possibility of a 
localized energy in general relativity, primarily due to the non-tensorial 
aspect of the pseudotensor, but others such as Bondi \cite{bondinew} argued that 
energy must be localizable. Moreover, even the deniers tacitly admit a certain 
degree of localization in any case as they speak of a material system 
\textit{losing} energy and that this loss is due to the gravity waves carrying 
away this energy. In other words, there is an energy to speak of \textit{within} 
the material source as well as \textit{within} the waves. 
  
2. \textit{What is the energy of the system during the particularly interesting 
gravity-wave emitting phase?}

 The standard derivations of the Tolman integral expression 
for energy break down when the metric 
is time-dependent. Therefore the energy at that point is unknown. In fact it is 
well to 
consider what criteria one would employ to decide upon the correctness of any 
energy
expression candidate in the dynamic state. There very 
well may be an additional part beyond the Tolman integral to express the energy
for a dynamical system that is a localized 
expression whose density involves the accelerations of the matter. If so, it 
would not always have a non-zero integral: the Tolman integral alone describes 
the total energy for a stationary source such as a rotating disk whose elements, by virtue of rotation,  
do accelerate. A useful goal for future research would be to determine if there 
is an additional part and if so, to deduce its value. 
 
3. \textit{Is the radiation-reaction type of calculation as solidly based in 
general 
relativity as it is in electromagnetism?}

 We regard this as an issue because Maxwell theory is linear and general 
relativity is non-linear. The radiation reaction in Maxwell theory (at least 
time-averaged) is meaningfully related to the unambiguous Poynting vector flux 
of electromagnetic radiation energy over even a lengthy period of time, in part 
because of the linearity. However, in the analogous problem in general 
relativity, there is no assurance that such a calculation can yield a meaningful 
result for anything other than a possibly very short time period beyond 
stationarity during which any non-linearities have not had sufficient time to 
grow to a significant size. Thus, while Eddington \cite{edd} argued that 
$v^2 \gg m/r$ for his spinning rod was a sufficient condition for a linearized 
treatment of the energy problem, the validity of linearization might be more 
severely restricted than he had believed. This has further ramifications, as we 
now discuss.

One might have argued that given the large reservoir of kinetic energy in a 
rapidly rotating Eddington rod as compared to the relatively small amount of 
gravitational contribution to the energy, and given the computed non-zero loss rate of the former 
(at least within the context of the assumed validity of linearization), 
it is inevitable that over a long period of time, the loss would accumulate to a 
large net energy loss value that could never be compensated by any gravitational 
component. However, question 3 above is relevant here. It is possible that the 
kinetic 
energy loss as computed traditionally holds only for a relatively short period 
beyond stationarity and that during this phase, there still remains an 
equalizing \textit{dynamical} gravitational component to render the localized 
energy conserved within the source. Moreover, as the non-linearities grow, it is 
possible that the kinetic energy loss component levels out and the vast 
kinetic energy reservoir can never really be tapped. Thus, an experimental test of the 
localization 
hypothesis would be one of actually observing the equivalent of a rapidly 
rotating \textit{strongly bound} (i.e. with $v^2 \gg m/r$) system such as an 
Eddington rod. The rate of change in 
period would have to decrease with the system approaching a state of uniform 
rotation as in classical physics for the viability of the hypothesis. On the 
other hand, if the decay should continue unabated, ultimately draining away the 
vast reservoir of kinetic energy, then the traditional view of an energy loss 
carried by the gravitational waves would be supported. Unfortunately, such 
strongly bound systems of sufficient strength to yield detectable results would 
be very hard to come by and the far more complex 
\textit{gravitationally} bound system of a rotating binary such as PSR 1913+16 
would not furnish such a test since in such a case, the store of gravitational 
energy is necessarily comparable to that of the kinetic energy store, i.e. $v^2$ 
of the order of $m/r$. 

We are presently exploring a problem that will hopefully by-pass the 
difficulties inherent
in the Eddington source. It is based upon an axially symmetric source that 
evolves from a \textit{static}
configuration, evolves into a dynamic phase with an imposed variation in its 
energy-momentum tensor
and then returns after a short time interval to a final rest configuration at 
lowest order. 
The goal is to integrate the change in the Tolman
integral over the brief history of evolution to determine whether or not the active gravitational mass at the end equals the
value at the beginning. By this approach, one avoids the uncertainty associated 
with the potential
build-up of non-linearities. Moreover, by completing the dynamic phase, the 
change or constancy in the 
Tolman integral would answer the energy-loss question in a definitive manner.

\section{Appendix}\label{sect:appendix}

\subsection{Dirac-Delta Integration \label{appendix:diracdelta}}

The integral of
\eqn{
	\int_{-\epsilon}^{\epsilon} f(x,y,t) \delta(\ybar) \,dy
}
is evaluated using the well-known identity
$\delta(ay-b)=\frac{1}{|a|}\delta(y-\frac{b}{a})$.  We assume that $t$ is
in the neighhbourhood of $t=0$ and so for some $\epsilon>0$, the limits
of integrations will completely cover the Dirac-delta distribution.
Factoring the coefficient out and evaluating the integral results in
\eqn{
	\int_{-\epsilon}^{\epsilon} f(x,y,t) \frac{1}{|\cos \omega t|}
	\delta\Bigl(y-\frac{x \sin \omega t}{\cos \omega t}\Bigr) \,dy
	= \frac{f(x,x\tan \omega t, t)}{\cos \omega t}
}
where we have assumed that $|\cos \omega t| \equiv \cos \omega t$ in the
neighbourhood of $t=0$. Thus the rule for integrating
$\delta(-x \sin \omega t + y \cos \omega t)$ WRT $y$ is
to substitute $y=x\tan\omega t$ and divide the entire quantity by $\cos\omega 
t$.

\subsection{Single Time-Derivative and Integral Commutation}
\label{commutation1}

It is very impractical to integrate before setting time $t=0$; setting $t=0$
prematurely could remove a large number of terms that are required for 
integration.  The identity,
\eqnn{leftonederivative}{
	\Bigl\{ \prt{}{t} \int_{-a\cos\omega t}^{a\cos\omega t}
		f(x,t) dx \Bigr\}_{\bigl| t=0}
	=\int_{-a}^{a} \Bigl\{\prt{f(x',t)}{t}\Bigr\}_{\bigl| t=0} dx'
}
for an arbitrary function $f(x,t)$ is extremely useful. It allows one to bring
the time-derivative through the time-dependent integral. The proof of this
identity is as follows.  \footnote{Of course this definition of $F(x,t)$ is
not unique.  Also, note that we could have defined
$F(x,t)=\int_{0}^{x} f(x',t) dx'$ but we cannot be
sure that there will be no singularities in the domain $[0,x]$.}
Define a function $F(x,t)$ as
\eqnn{mydefF}{
	F(x,t) = \int_{0}^{x} f(x',t) dx'
}
with the condition that
$\partial{^2 F(x,t)}/{\partial x\partial t}
= \partial{^2 F(x,t)}/{\partial t\partial x}$.
Using this definition in the integral on the LHS of 
(\ref{leftonederivative}),
we find that
\eln{
	\int_{-a\cos\omega t}^{a\cos\omega t}  f(x,t) dx 
	&= F(x,t)_{\bigl|x=a\cos\omega t}-F(x,t)_{\bigl|x=-a\cos\omega t}\\
	&= F(a\cos\omega t,t) -  F(-a\cos\omega t,t).
}
When we apply the time-derivative, the chain rule is used on the
$x=\pm a\cos\omega t$ terms and the equation becomes
\elnn{firstderivativelemma}{
	\prt{}{t} \int_{-a\cos\omega t}^{a\cos\omega t}  f(x,t) dx 
	=& \Bigl( \prt{F(x,t)}{x} \frac{d}{dt}(a\cos\omega t) +
		\prt{F(x,t)}{t} \Bigr)_{\bigl| x=a\cos\omega t} \\
	&-\Bigl( \prt{F(x,t)}{x} \frac{d}{dt}(-a\cos\omega t) +
		\prt{F(x,t)}{t} \Bigr)_{\bigl| x=-a\cos\omega t} \\
	=&-a\omega\sin\omega t \, ( f(a\cos\omega t,t)+f(-a\cos\omega t,t))\\
	&+\Bigl(\prt{F(x,t)}{t} \Bigr)_{\bigl| x=a\cos\omega t} 
		-\Bigl(\prt{F(x,t)}{t} \Bigr)_{\bigl| x=-a\cos\omega t} 
}
The link, $x=a\cos\omega t$, is applied at the very last
step--{\em outside the parenthesis}.  Hence, there is no need to be concerned 
about
time-varying integral limits when we do take the time-derivative
of (\ref{mydefF}).  Now, once we set $t=0$, we find that the $\sin\omega t$
terms go to zero simplifying the equation,
\eln{
	\Bigl\{ \prt{}{t} \int_{-a\cos\omega t}^{a\cos\omega t}
		f(x,t) dx \Bigr\}_{\bigl| t=0}
	=& \Bigl( 0 +  \prt{F(x,t)}{t} \Bigr)_{\bigl| x=a,t=0} \\
	&-\Bigl( 0 +  \prt{F(x,t)}{t} \Bigr)_{\bigl| x=-a,t=0}.
}
To show that this is indeed the RHS of (\ref{leftonederivative}),
we refer back to the definition of $F(x,t)$ in (\ref{mydefF}).
The expression,
\eqn{
	\Bigl( \prt{F(x,t)}{t} \Bigr)_{\bigl| x=a,t=0} 
		-\Bigl( \prt{F(x,t)}{t} \Bigr)_{\bigl| x=-a,t=0}
	=
	\Bigl( \prt{}{t} \int_{-a}^{a} f(x',t) dx' \Bigr)_{\bigl| t=0}
}
is equivalent to the RHS of (\ref{leftonederivative}) because we can now move
the time-derivative through the integral.

\subsection{Double Time-Derivative and Integral Commutation}
With a similar approach we can find an expression for the second time-derivative
as we now demonstrate:
\eqneqnn{secondderivativelemma}{
	\Bigl\{ \prt{^2}{t^2} \int_{-a\cos\omega t}^{a\cos\omega t}
		f(x,t) \,dx \Bigr\}_{\bigl|t=0} \\
	= -a\omega^2 ( f(a,0)+f(-a,0) )
	+\int_{-a}^{a} \Bigl\{\prt{^2 f(x',t)}{t^2} \Bigr\}_{\bigl|t=0} \,dx'.
}
To prove this, we first define an intermediary function,
\eqnn{mydefG}{
	G(x,t) = \prt{F(x,t)}{t}.
}
When we apply another time-derivative to (\ref{firstderivativelemma}) with the
help of (\ref{mydefG}), it becomes
\eln{
	\prt{^2}{t^2} \int_{-a\cos\omega t}^{a\cos\omega t}  f(x,t) \,dx 
	=& \prt{}{t} \Bigl(
	-a\omega\sin\omega t\,(f(a\cos\omega t,t)+f(-a\cos\omega t,t))\Bigr) \\
	&+\prt{}{t} \Bigl(
	G(x,t)_{\bigl| x=a\cos\omega t}-G(x,t)_{\bigl| x=-a\cos\omega t}\Bigr).
}
We will not need to expand the first two terms completely because when we
set $t=0$, many parts of
\eln{
	\prt{^2}{t^2} \int_{-a\cos\omega t}^{a\cos\omega t}  f(x,t) dx 
	= &
	-a\omega^2\cos\omega t\Bigl(f(a\cos\omega t,t)
	+f(-a\cos\omega t,t)\Bigr)\\
	&-a\omega\sin\omega t  \prt{}{t} \Bigl( f(a\cos\omega t,t)
		+f(-a\cos\omega t,t)  \Bigr) \\
	&+ \Bigl(\prt{G(x,t)}{x} \frac{d}{dt} (a\cos\omega t) 
		+ \prt{G(x,t)}{t} \Bigr)_{\bigl| x=a\cos\omega t} \\
	&- \Bigl(\prt{G(x,t)}{x} \frac{d}{dt} (-a\cos\omega t) 
		+ \prt{G(x,t)}{t} \Bigr)_{\bigl| x=-a\cos\omega t}
}
will cancel. With $t=0$, the expression simplifies to
\eln{
	\Bigl\{ \prt{^2}{t^2} \int_{-a\cos\omega t}^{a\cos\omega t}
		f(x,t) \,dx \Bigr\}_{\bigl|t=0}
	= & -a\omega^2\Bigl( f(a,0)+f(-a,0) \Bigr) \\
	&+ \Bigl(\prt{G(x,t)}{t} \Bigr)_{\bigl| x=a, t=0 }
	- \Bigl( \prt{G(x,t)}{t} \Bigr)_{\bigl| x=-a, t=0}.
}
Based on the definition for $G(x,t)$ in (\ref{mydefG}), this is
\eqneqn{
	\Bigl\{ \prt{^2}{t^2} \int_{-a\cos\omega t}^{a\cos\omega t}
		f(x,t) \,dx \Bigr\}_{\bigl|t=0}
	= 
	-a\omega^2\Bigl( f(a,0)+f(-a,0) \Bigr) \\
	+ \Bigl(\prt{^2 F(x,t)}{t^2} \Bigr)_{\bigl| x=a, t=0 }
	- \Bigl( \prt{^2 F(x,t)}{t^2} \Bigr)_{\bigl| x=-a, t=0}.
}
And finally, using (\ref{mydefF}), we can see that this 
\eqneqn{
	-a\omega^2\Bigl( f(a,0)+f(-a,0) \Bigr) 
	+ \Bigl(\prt{^2 F(x,t)}{t^2} \Bigr)_{\bigl| x=a, t=0 }
	- \Bigl( \prt{^2 F(x,t)}{t^2} \Bigr)_{\bigl| x=-a, t=0} \\
	=
	-a\omega^2\Bigl( f(a,0)+f(-a,0) \Bigr) 
	+\Bigl\{ \prt{^2}{t^2} \int_{-a}^{a} f(x',t)dx'\Bigr\}_{\bigl|t=0}
}
is equivalent to the RHS of (\ref{secondderivativelemma}).

\subsection{Single Time-Derivative of Step-Function}

Another proof of section \ref{commutation1} can be done using step-functions.
The basic reason why there is time-dependence in the integral limits
of (\ref{leftonederivative}) is because of the step-functions in the
integrand. An equivalent statement to (\ref{leftonederivative}) is
\eqneqnn{stepfunctionclaim1}{
	\Big\{ \prt{}{t} \int_{-a}^a \int_{-\epsilon}^\epsilon 
	(H(a-\bar x)+H(a+\bar x)-1) g(x,y,t) \delta(\bar y) \,dy dx
	\Bigr\}_{\bigl| t=0} \\
	=
	\int_{-a}^a \int_{-\epsilon}^\epsilon 
	\Big\{
	 \prt{}{t}\Bigl( g(x,y,t) \delta(\bar y) \Bigr)
	\Bigr\}_{\bigl| t=0}
	\,dy dx
}
with the relationship between (\ref{leftonederivative})
and (\ref{stepfunctionclaim1}) being
$f(x,t)
	=\int_{-\epsilon}^\epsilon  g(x,y,t) \delta(\bar y) \,dy
	= g(x,x\tan \omega t,t)/\cos\omega t$.
We can move the time derivative through the integral
of (\ref{stepfunctionclaim1}), yielding
\eqneqnn{d1step}{
	\prt{}{t} \int_{-a}^a \int_{-\epsilon}^\epsilon 
	(H(a-\bar x)+H(a+\bar x)-1) g(x,y,t) \delta(\bar y) \,dy dx \\
	=\int_{-a}^a \int_{-\epsilon}^\epsilon 
	(-\omega\bar y\delta(\bar x - a) 
	+\omega\bar y \delta(\bar x + a)) g(x,y,t)\delta(\bar y) \\
	+(H(a-\bar x)+H(a+\bar x)-1) 
		\prt{}{t}\Bigl( g(x,y,t)\delta(\bar y) \Bigr) \,dy dx.
}
Having now dispensed with the derivative apart from in the very last term, we 
may set $t=0$.
This implies that $\bar x = x$, $\bar y\rightarrow 0$, and
$H(a\pm\bar x)\equiv 1$. Thus,
\eqneqn{
	\Big\{ \prt{}{t} \int_{-a}^a \int_{-\epsilon}^\epsilon 
	(H(a-\bar x)+H(a+\bar x)-1) g(x,y,t) \delta(\bar y) \,dy dx
	\Bigr\}_{\bigl| t=0} \\
	=\int_{-a}^a \int_{-\epsilon}^\epsilon 
	(-\omega y\delta(y)\delta(x-a)
		+\omega y\delta(y)\delta(x+a))g(x,y,0) \\
	+(1) \prt{}{t}\Bigl( g(x,y,t)\delta(\bar y)
		\Bigr)_{\bigl| t=0} \,dy dx.
}
Using the fact that $y\delta(y)\equiv 0$, the first two terms vanish,
leading us to (\ref{stepfunctionclaim1}).

\subsection{Double Time-Derivative of Step-Function}

The equivalent statement to (\ref{secondderivativelemma}) using step
functions is
\eqneqnn{stepfunctionclaim2}{
	\Big\{ \prt{^2}{t^2} \int_{-a}^a \int_{-\epsilon}^\epsilon 
	(H(a-\bar x)+H(a+\bar x)-1) g(x,y,t) \delta(\bar y) \,dy dx
	\Bigr\}_{\bigl| t=0} \\
	=
	-a\omega^2\Bigl( g(a,0,0)+g(-a,0,0) \Bigr)
	+\int_{-a}^a \int_{-\epsilon}^\epsilon 
	\Big\{
	 \prt{^2}{t^2}\Bigl( g(x,y,t) \delta(\bar y) \Bigr)
	\Bigr\}_{\bigl| t=0}
	\,dy dx.
}
To prove this, we first apply another time-derivative  to  (\ref{d1step})
to get
\eqneqn{
	\prt{^2}{t^2} \int_{-a}^a \int_{-\epsilon}^\epsilon 
	(H(a-\bar x)+H(a+\bar x)-1) g(x,y,t) \delta(\bar y) \,dy dx \\
	=\int_{-a}^a \int_{-\epsilon}^\epsilon 
	\Bigl(\omega^2\bar x\delta(\bar x - a)
		+\omega^2\bar y^2 \frac{\delta(\bar x - a)}{(a-\bar x)}
	-\omega^2\bar x\delta(\bar x + a)
		-\omega^2\bar y^2 \frac{\delta(\bar x + a)}{-(\bar x + a)}
	\Bigr) g(x,y,t)\delta(\bar y) \\
	+2\Bigl(-\omega\delta(\bar x - a)\bar y 
		+\omega\delta(\bar x + a)\bar y  \Bigr)
	 \Bigl(  \delta(\bar y)\prt{g(x,y,t)}{t} 
		+\frac{\omega\bar x}{\bar y}\delta(\bar y)g(x,y,t)\Bigr) \\
	+(H(a-\bar x)+H(a+\bar x)-1)
		\prt{^2}{t^2}\Bigl(g(x,y,t)\delta(\bar y)\Bigr)\,dy dx.
}
Now, set $t=0$ and we get
\eqneqn{
	\Big\{ \prt{^2}{t^2} \int_{-a}^a \int_{-\epsilon}^\epsilon 
	(H(a-\bar x)+H(a+\bar x)-1) g(x,y,t) \delta(\bar y) \,dy dx
	\Bigr\}_{\bigl| t=0} \\
	=\int_{-a}^a \int_{-\epsilon}^\epsilon 
	\Bigl(\omega^2x\delta(x - a) -\omega^2 x\delta( x + a) \Bigr)
		g(x,y,0)\delta( y) \\
	+2\Bigl(-\omega\delta( x - a) y +\omega\delta( x + a) y \Bigr)
	 \frac{\omega x}{ y}\delta( y) g(x,y,0) \\
	+(1)\Bigl\{\prt{^2}{t^2}\Bigl( g(x,y,t)\delta(\bar y)
		\Bigr)\Bigr\}_{\bigl| t=0} \,dy dx 
}
The first four terms of the RHS collapse to
$-a\omega^2 ( g(a,0,0)+g(-a,0,0) )$ and hence the whole expression,
\eqn{
	-a\omega^2 ( g(a,0,0)+g(-a,0,0) )
	+\int_{-a}^a \int_{-\epsilon}^\epsilon 
	\Bigl\{\prt{^2}{t^2}\Bigl( g(x,y,t)\delta(\bar y)
	\Bigr)\Bigr\}_{\bigl| t=0} \,dy dx,
}
is equivalent of the RHS of (\ref{stepfunctionclaim2})

-----------------------------------------------------------------------------

Acknowledgment: This work was supported by grants from the Natural 
Sciences and Engineering Research Council of Canada and the University of 
Victoria Faculty Research Fund. We thank M. Santoprete for his kind assistance.
         

\begin{thebibliography}{99}
\bibitem{conventions}
Latin indices range 0-3, Greek indices range 1-3, G is the gravitational 
constant, c is the speed of light. Where convenient, G and c are set to 1.
\end{thebibliography}
\end{document}